\documentclass[11pt,a4paper,useAMS,usenatbib]{emulateapj}
\usepackage{epsfig}
\usepackage{longtable}
\usepackage{comment}
\usepackage{color}
\usepackage[normalem]{ulem}
\usepackage[colorlinks=true, pdfstartview=FitV, linkcolor=blue,citecolor=blue, urlcolor=blue]{hyperref}
\usepackage{amsmath}
\usepackage{natbib}

\newcommand{\etal}{et al.}

\shorttitle{Bow Shock in the 3C 438 Cluster}
\shortauthors{EMERY ET AL.}

\begin{document}

\title{A Spectacular Bow Shock in the 11 keV Galaxy Cluster Around 3C 438}

\author{Deanna L. Emery\altaffilmark{1},
\'Akos Bogd\'an\altaffilmark{1}, 
Ralph P. Kraft\altaffilmark{1}, 
Felipe Andrade-Santos\altaffilmark{1},
William R. Forman\altaffilmark{1},
Martin J. Hardcastle\altaffilmark{2}, and 
Christine Jones\altaffilmark{1}}
\altaffiltext{1}{Harvard-Smithsonian Center for Astrophysics\\ 60 Garden St., Cambridge, MA 02138\\ E-mail: abogdan@cfa.harvard.edu}
\altaffiltext{2}{School of Physics, Astronomy and Mathematics, University of Hertfordshire, College Lane, Hatfield, Hertfordshire AL10 9AB, UK}

\begin{abstract}

We present results of deep 153 ks Chandra observations of the hot, 11 keV, galaxy cluster associated with the radio galaxy 3C 438. By mapping the morphology of the hot gas and analyzing its surface brightness and temperature distributions, we demonstrate the presence of a merger bow shock. We identify the presence of two jumps in surface brightness and in density located at $\sim$400 kpc and $\sim$800 kpc from the cluster's core. At the position of the inner jump, we detect a factor of $2.3\pm 0.2$ density jump, while at the location of the outer jump, we detect a density drop of a factor of $3.5 \pm 0.7$. Combining this with the temperature distribution within the cluster, we establish that the pressure of the hot gas is continuous at the 400 kpc jump, while there is a factor of $6.2 \pm 2.8$ pressure discontinuity at 800 kpc jump. From the magnitude of the outer pressure discontinuity, using the Rankine-Hugoniot jump conditions, we determine that the sub-cluster is moving at $M = 2.3\pm 0.5$, or approximately $2600\pm 565$ km/s through the surrounding intracluster medium, creating the conditions for a bow shock. Based on these findings, we conclude that the pressure discontinuity is likely the result of an ongoing major merger between two massive clusters. Since few observations of bow shocks in clusters have been made, this detection can contribute to the study of the dynamics of cluster mergers, which offers insight on how the most massive clusters may have formed.

\end{abstract}


\section{Introduction}

Galaxy clusters -- the largest bound structures in the Universe -- are believed to grow hierarchically via mergers of galaxy groups and smaller sub-clusters and from accretion of gas and dark matter \citep[see][and references therein]{Kravtsov2012}. The energy from mergers is largely dissipated in the gas within the cluster and often leaves an imprint on the distribution of the hot intracluster medium (ICM). Indeed, X-ray observations of galaxy clusters pointed out the presence of shocks and cold fronts \citep[e.g.][]{Vikhlinin2001,Markevitch2002,Ascasibar2006,Markevitch2007}, and signatures of ram-pressure stripping \citep{Nulsen1982,Close2013}. Studies of such disturbances provide a key laboratory to study the gas dynamics within an evolving system. Major mergers of galaxy clusters are particularly important, since these relatively rare and energetic events have the most dramatic, long-lasting effects on the cluster gas, and hence play a key role in the energy budget and thermodynamic evolution of galaxy clusters. 

The galaxy cluster, associated with the powerful radio galaxy 3C 438, is included in the \textit{Planck} 2015 release \citep{Planck2015}. The signal-to-noise ratio of this detection is $S/N=11.07$, making this cluster among the top $10\%$ of the most significant detections in Planck's 1653 total cluster detections. This is confirmed by the mass estimate obtained from the $Y_{\rm SZ} - M_{\rm tot}$ scaling relation, which estimates the total mass of the galaxy cluster to be $7.83^{+0.43}_{-0.44} \times 10^{14} \rm{M_{\odot}}$.

The galaxy cluster around 3C 438 has been previously observed by the \textit{Chandra} X-ray Observatory \citep{Kraft2007}. The short 30 ks observation demonstrated the extremely hot ICM temperature, which made it one of the hottest clusters, rivaling the ``Bullet'' cluster \citep{Markevitch2002}. These data revealed surface brightness discontinuities extending more than 550 kpc ($2.67\arcmin$). To explain the origin of the complex morphology, \citet{Kraft2007} suggested that it may either arise from the result of a radio outburst or an on-going major merger. However, these scenarios could not be differentiated based on the previous X-ray observation.

\begin{figure*}
\centering
\includegraphics[width=16cm]{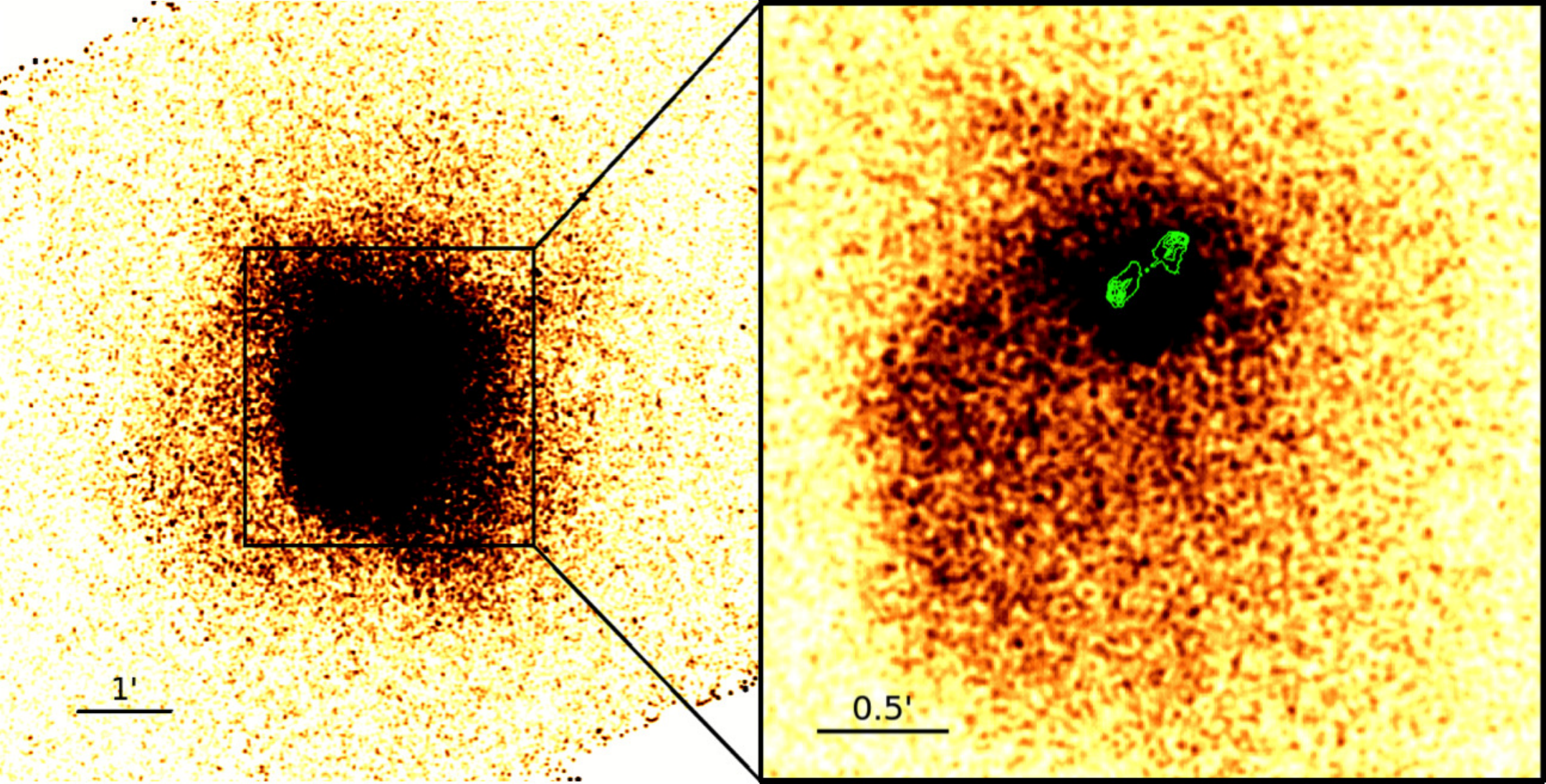}
\caption{An image of the 3C 438 cluster filtered for the 0.5-4 keV energy range. The image was smoothed with a Gaussian kernel of three-pixel size. At $z=0.29$, one arcminute corresponds to 261 kpc. Contours of the radio lobes of the radio galaxy 3C 438 from a VLA observation at 1.5 GHz \citep{Harwood2015} are overlaid in the right panel, and are approximately 45 kpc in size, which corresponds to $\sim 0.17\arcmin$ ($\sim 10\arcsec$). Note that north corresponds to the top of the image, and east corresponds to the left of the image. The morphology of the hot gas in the cluster appears to be strongly disturbed, which indicates that it has undergone a recent interaction. There are two notable surface brightness discontinuities on the eastern side of the cluster (see also Figure \ref{fig:outlines}). The first is located at approximately $1.5\arcmin$, or $\sim 400$ kpc, from the center of the cluster (left panel), and the second is located at approximately $3\arcmin$, or $\sim 800$ kpc, from the center (right panel).}
\label{fig:image}
\end{figure*}

In this paper, we study follow-up \textit{Chandra} observations of the galaxy cluster associated with 3C 438. Our main goal is to determine the cause of the large-scale surface brightness features. This paper is structured as follows: \ref{sec:analysis} discusses the data and reductions. In Section \ref{sec:results}, we describe the methods used in the analysis of the data and present the results, including the surface brightness profiles, temperature profile, and density profile. In Section \ref{sec:discussion}, we discuss our conclusions concerning the morphology of the cluster. Finally, in Section \ref{sec:conclusions}, we summarize the key results and suggest future lines of investigation. Throughout this paper, we assume standard $\Lambda$CDM cosmology with $H_0 = 70 \ \rm{km \ s^{-1}} \ Mpc^{-1}$, $\Omega = 0.3$, and $\Omega_{\Lambda}=0.7$. The redshift of 3C 438 is $z = 0.29$, which implies a linear scale of 261 kpc arcmin$^{-1}$ and a luminosity distance of 1493.2 Mpc. The galaxy cluster lies at low Galactic latitude and has a correspondingly large line-of-sight column density. Based on the IRAS $100 \ \mu$m dust map and using the relation between hydrogen column density ($N_{\rm H}$) and optical extinction ($A_{\rm V}$) we computed $N_{\rm H} =2.6\times10^{21} \ \rm{cm^{-2}}$ \citep{Guver2009}.

\section{Data analysis}
\label{sec:analysis}

\textit{Chandra} has observed 3C 438 and the associated galaxy cluster three times for a total of 166.9 ks. Further details about each observation are listed in Table \ref{table:observations}. The reduction and analysis of the data were performed using standard CIAO\footnote{http://cxc.harvard.edu/ciao/} software package tools (version 4.7) and the associated CALDB (version 4.6.9) \citep{Fruscione2006}.

\begin{table}
\begin{center}
\caption{\textit{Chandra} Observations used in the analysis.\label{tbl-1}}
\begin{tabular}{crrrrr}
\tableline
\tableline
Obs-ID & $t_{\rm total}$ (ks)& $t_{\rm net}$ (ks)& Instrument & Date \\
\tableline
3967   & 47.5  & 34.1  & ACIS-S & 2002 Dec 27 \\
12879 & 72.0  & 71.8  & ACIS-S & 2011 Jan 30 \\
13218 & 47.4  & 47.4  & ACIS-S & 2011 Jan 28 \\
\tableline
\label{table:observations}
\end{tabular}
\end{center}
\end{table}

We reprocessed all data sets, which assures that the newest calibration updates are applied. As our main science goal is to study the ICM around 3C 438, we identified and masked out bright point sources. To detect point sources, which are mostly background AGN, we used the CIAO \textsc{wavdetect} tool and searched for point sources in the  $0.5-7$ keV energy range. We optimize the efficiency of the source detection algorithm by searching for sources on the wavelet scale sizes of 1.0, 1.414, 2, 2.828, 4, 5.567, 8, and by modifying the {\it ellsigma} parameter to 4.0. The detected point sources were then masked out from all further analysis of the ICM. 

The observations were then individually filtered for flare-contaminated intervals. To do this, we extracted light curves in the $2.3-7.3$ keV energy range since \textit{Chandra} is most sensitive to Solar flares at these energies \citep{hickox2006}. We then binned the light curves into 200-second bins, and computed the mean count rate on the ACIS-S3 detector. We removed all time bins where the count rates differed from the mean by $\pm 3$ sigma. Only one observation, Obs-ID 3967, was slightly flare contaminated, with about 30\% of the exposure time excluded. The resulting total clean exposure time is 153.3 ks. 

To account for vignetting effects, we created weighted exposure maps for each observation, filtered for an energy range of 0.5 to 4 keV. The exposure maps were made assuming an optically-thin thermal plasma emission model (\textsc{apec} model in \textsc{XSpec}) with parameters of $N_{\rm H} = 2.7 \times 10^{21} \ \rm{cm^{-2}}$ and $kT=10$ keV, and abundance of $0.4$ Solar. This particular choice of spectrum was determined by the parameter values generated by a best-fit spectrum for a circular region centered on the cluster's core with a radius of $3\arcmin$.
The individual exposure maps were projected to the coordinate system of Obs-ID 12879 and were co-added for further analysis. 

The galaxy cluster associated with 3C 438 is located at low Galactic latitude and, thus, has a high line-of-sight column density, $N_{\rm H} = 2.7 \times 10^{21} \ \rm{cm^{-2}}$. Due to the high level of soft Galactic emission, the ACIS blank-sky files provided in the Chandra Calibration Database cannot accurately account for the background components. Since the blank-sky dataset has been produced using a set of \textit{Chandra} observations at high Galactic latitude, by utilizing these background data, the sky background components would be over-subtracted at energies below approximately 2 keV. For this reason, we opted to use local regions to account for the sky and instrumental background components. When creating the background files, the fact that the cluster extends onto two detectors, ACIS-S2 and ACIS-S3, had to be taken into account. To this end, we used the local background level on ACIS-S3 to account for the background on this detector, while we used the average background levels of ACIS-I2 and ACIS-I3 to account for the background level on ACIS-S2. For each case, the background levels were re-normalized using the count rate ratios observed in the $10-12$ keV band.


\begin{figure}
\centering
\includegraphics[width=8cm]{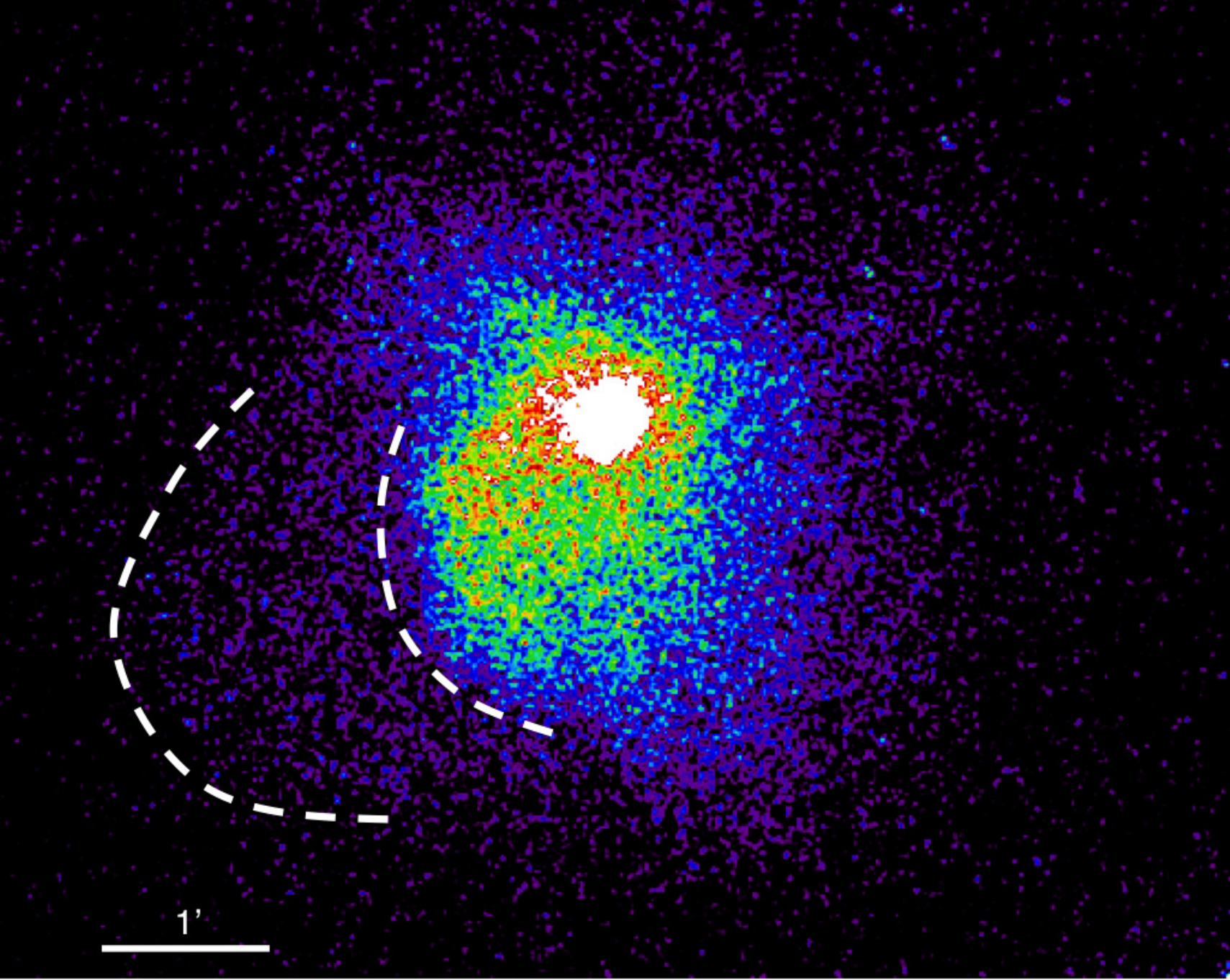}
\caption{The two surface brightness discontinuities are highlighted by the green curves. The inner discontinuity is a cold front located at $1.5\arcmin$ (400 kpc) from the cluster's core. The outer discontinuity is a shock front located at $3\arcmin$, or 800 kpc from the center of the cluster. The surface brightness of the outer discontinuity exhibits an angular shape similar to a Mach cone, as opposed to a circular shape centered on the cluster core.}
\label{fig:outlines}
\end{figure}

\section{Results}
\label{sec:results}

\subsection{Images}
In Figure \ref{fig:image} we present merged $0.5-4$ keV band \textit{Chandra} X-ray images of the core (left panel) and the large scale surroundings (right panel) of the galaxy cluster. Both images are smoothed with a Gaussian with a kernel size of three pixels. Bright resolved point sources are excluded and for illustration purposes, we filled their former locations with the level of the surrounding X-ray emission using the CIAO \textsc{dmfilth} tool. 
These images demonstrate significant X-ray emission originating from the optically-thin hot gas with $kT \sim10$ keV (see Section \ref{sec:profiles}). The X-ray images reveal the complex distribution of the hot gas both in the core and at large scales, suggesting that the cluster has not yet settled into a relaxed state. The most striking feature is the surface brightness discontinuity located at about $1.5\arcmin$ ($\sim400$ kpc) from the cluster core, which extends more than $2.67\arcmin$ (550 kpc). At even larger scales -- at about $3\arcmin$ ($\sim800$ kpc) from the cluster core -- another surface brightness discontinuity is present. 

Overall, these discontinuities suggest that the cluster is highly disturbed on large scales, and the features may be the result of a recent interaction or merger. The presence of two sharp surface brightness discontinuities is similar to that observed in the ``Bullet cluster,'' which is the prototypical example of two clusters undergoing a supersonic merger \citep{Markevitch2002}. To constrain the origin of the sharp surface brightness features, we measure the temperature and pressure jumps across the surface brightness discontinuities. This will allow us to probe whether the inner surface brightness jump originates from a supersonic infall and if the outer discontinuity can be attributed to a bow shock, or whether this is due to some large scale subsonic motions such as sloshing \citep{Ascasibar2006}.

\begin{figure*}
\centering
\includegraphics[width=8cm]{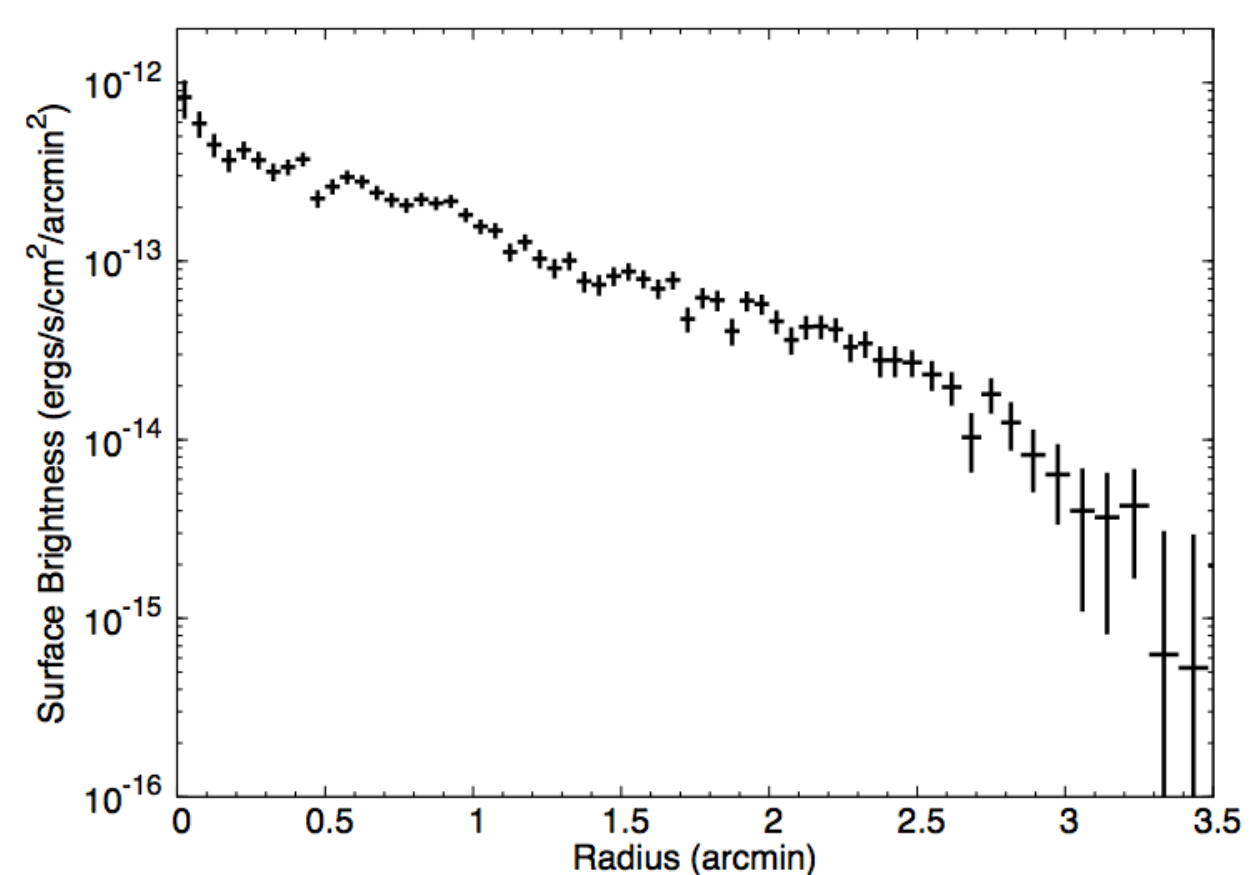}
\includegraphics[width=8cm]{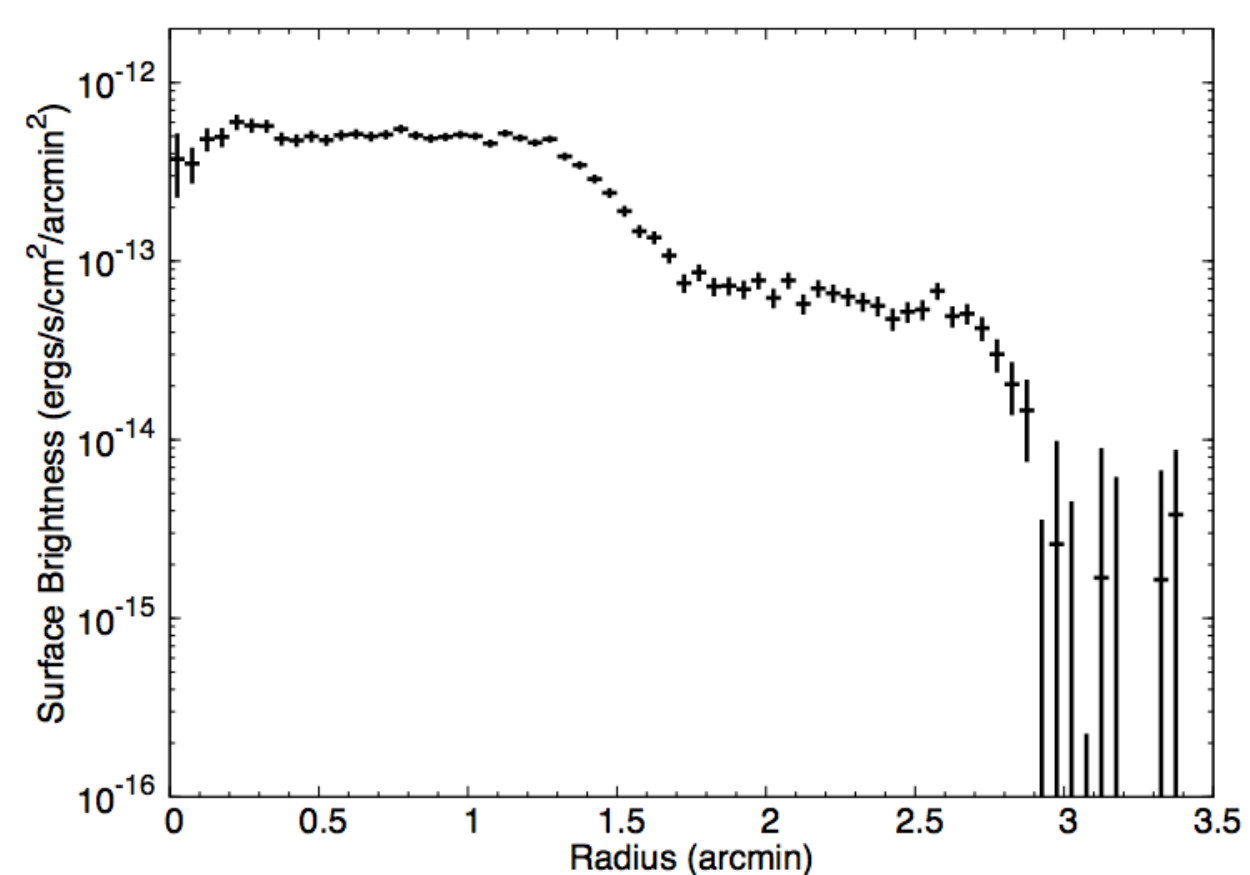}
\caption{The surface brightness profile on the right corresponds to the wedge extending $3.5\arcmin$ to the north-west from the center of the cluster. Each bin is $0.05\arcmin$ in size. The surface brightness profile on the left corresponds to the region that extends $3.5\arcmin$ to the east from the center of the cluster. Here again, each bin has a width of $0.05\arcmin$.}
\label{fig:sb}
\end{figure*}

\begin{figure}
\centering
\includegraphics[width=8cm]{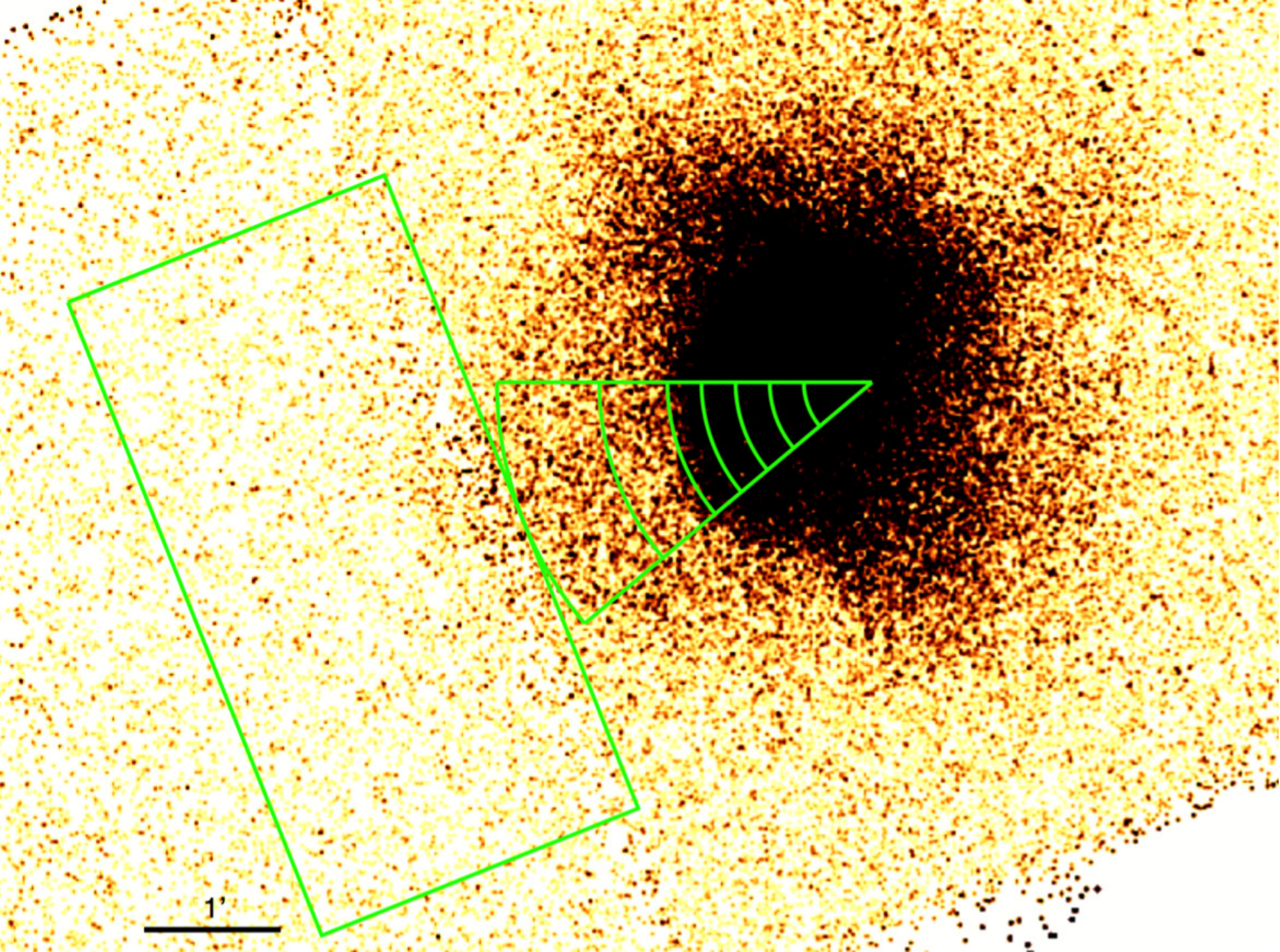}
\caption{The regions shown in this figure were used to construct the temperature profile shown in Figure \ref{fig:temperature}. The circular wedges extend $2.75\arcmin$ to the east and the box has a size of $2.5\arcmin \times 5\arcmin$. Note that the region is not centered on the core of the cluster.} 
\label{fig:temp_regions}
\end{figure}

\subsection{Profiles}
\label{sec:profiles}
We create surface brightness, density, and temperature profiles to quantify the changes across the edges. The density and temperature profiles together will allow us to derive the pressure jump across the edges, and hence reveal the origin of the discontinuities. Since the cluster gas exhibits high temperatures, we extract the surface brightness and density profiles from the $0.5-4$ keV band X-ray images, thereby maximizing the signal-to-noise ratio. 

In Figure \ref{fig:sb}, we plot the background-subtracted, vignetting-corrected surface brightness profiles of the cluster gas. To extract the profiles, we utilized circular wedges towards the east and the west with position angles of $20\degr-  60\degr$ and $180\degr-  220\degr$ respectively. Since we aimed to align the position of the regions with the sharp discontinuities in the hot gas distribution, the regions are centered south of the cluster's core (RA = 21:55:50.65, DEC =  +37:59:57.11). The surface brightness profiles confirm the picture hinted by the images. The profile towards the east demonstrates two distinct jumps at radii of $\sim1.5\arcmin$ and $\sim3\arcmin$, where the surface brightness drops by a factor of about 10. As opposed to this, the profile towards the west does not reveal any significant surface brightness jumps, and instead exhibits a gradual decrease.

Although the projected surface brightness profile points out the approximate locations of the discontinuities, due to projection effects, their exact positions relative to the center of mass cannot be determined from these profiles. To constrain the positions of the gas density jumps and determine their magnitudes, we perform de-projection analysis and build de-projected density profiles. To this end, we fit the inner discontinuity by describing the surface brightness profile within each wedge assuming spherical symmetry and isothermal temperature for the gas. We assume that the gas density follows a $\beta$-model inside the discontinuity and a power law model outside the edge. For the outer jump we used two power law models, where the parameters of the first power law were fixed at the values obtained from those of the inner jump. These models are described mathematically as: 
\begin{eqnarray}
n(r)=\left\{ \begin{array}{ll}
\renewcommand{\arraystretch}{3}
A\left[1 + (r/r_{\rm c})^2\right]^{-3\beta/2},
            & \mbox{\hspace{0.9cm} $r \leq r_{\rm cut}$}\\
B \left(r/r_{\rm c}\right)^{-\alpha},
            & \mbox{\hspace{0.9cm} $r > r_{\rm cut}$}\\
\end{array}
\right.
\label{eq:compression}
\end{eqnarray}
where $r_{\rm{c}}$ is the radius of the cluster core, and $r_{\rm cut}$ is the radius
at which a density jump occurs. Here, $A$ and $B$ are related by:
\begin{eqnarray}
B = \frac{A \left[1 + (r_{\rm cut}/r_{\rm
      c})^2\right]^{-3\beta/2}}{C \left(r_{\rm cut}/r_{\rm c}\right)^{-\alpha}}
\end{eqnarray}
where $C$ is the density jump. When describing the gas density of the inner discontinuity, it is important to realize that the surface brightness profile was not centered on the peak of the X-ray surface brightness, coincident with the Brightest Cluster Galaxy, the host galaxy of 3C 438. This results in a virtually flat density profile within the inner $\sim1\arcmin$ (Figure \ref{fig:sb} right panel). A direct consequence of fitting the flat density profile with a $\beta$-model is that $\beta$ and $r_{\rm{c}}$ parameters became degenerate. To avoid this difficulty, we fixed the value of the former parameter at $\beta=0.3$. It is possible that a second Brightest Cluster Galaxy is near the vertex of the defined sectors. However, we were unable to confirm the presence of such a galaxy due the absence of deep and large field-of-view images directed towards 3C 438.

\begin{figure}
\centering
\includegraphics[scale=0.65]{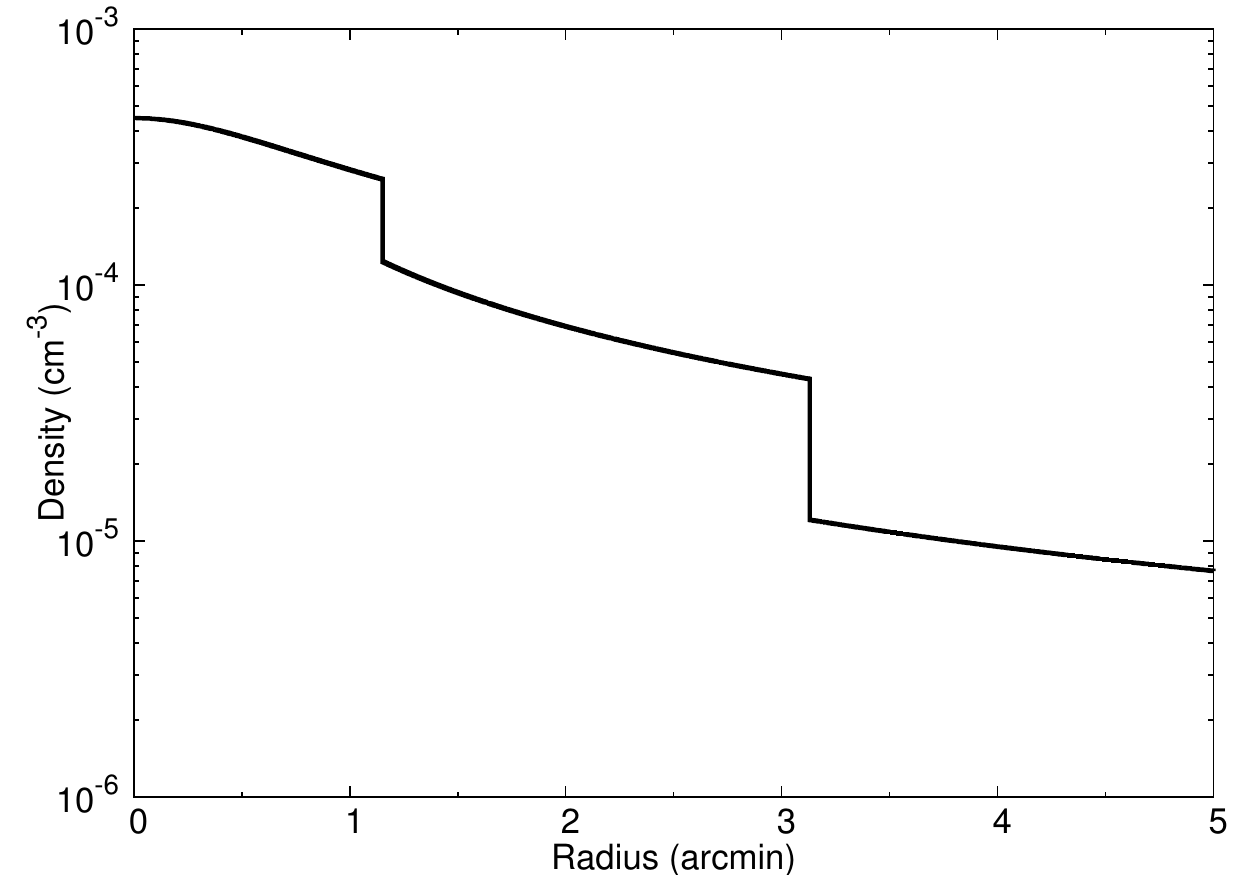}
\caption{The deprojected density profile obtained by the Proffit software was created from a best-fit for a density model consisting of a $\beta$-model and two power laws. The plot shows a drop in density by a factor of $2.3\pm 0.2$ at $1.5\arcmin$, and a drop by a factor of $3.5 \pm 0.7$ at $3\arcmin$.}
\label{fig:deproject}
\end{figure}

By utilizing the \textsc{proffit} software package \citep{Eckert2011}, we fit the de-projected density profiles obtained from the $0.5-4$ keV band images. The inner and outer jumps were fit using the regions between $1\arcmin-2.7\arcmin$ and $2.5\arcmin-5.0\arcmin$, respectively. By fitting the density profiles with the above described models, we obtained a density drop of a factor of $2.33\pm 0.15$ at $1.53\arcmin$ and $3.55\pm 0.76$ at $3.13\arcmin$. By combining the best-fit parameters for the density profile, we build the de-projected gas density profile in Figure \ref{fig:deproject}. The best-fit parameters are given in Table 2. 

\begin{figure}
\centering
\includegraphics[width=8cm]{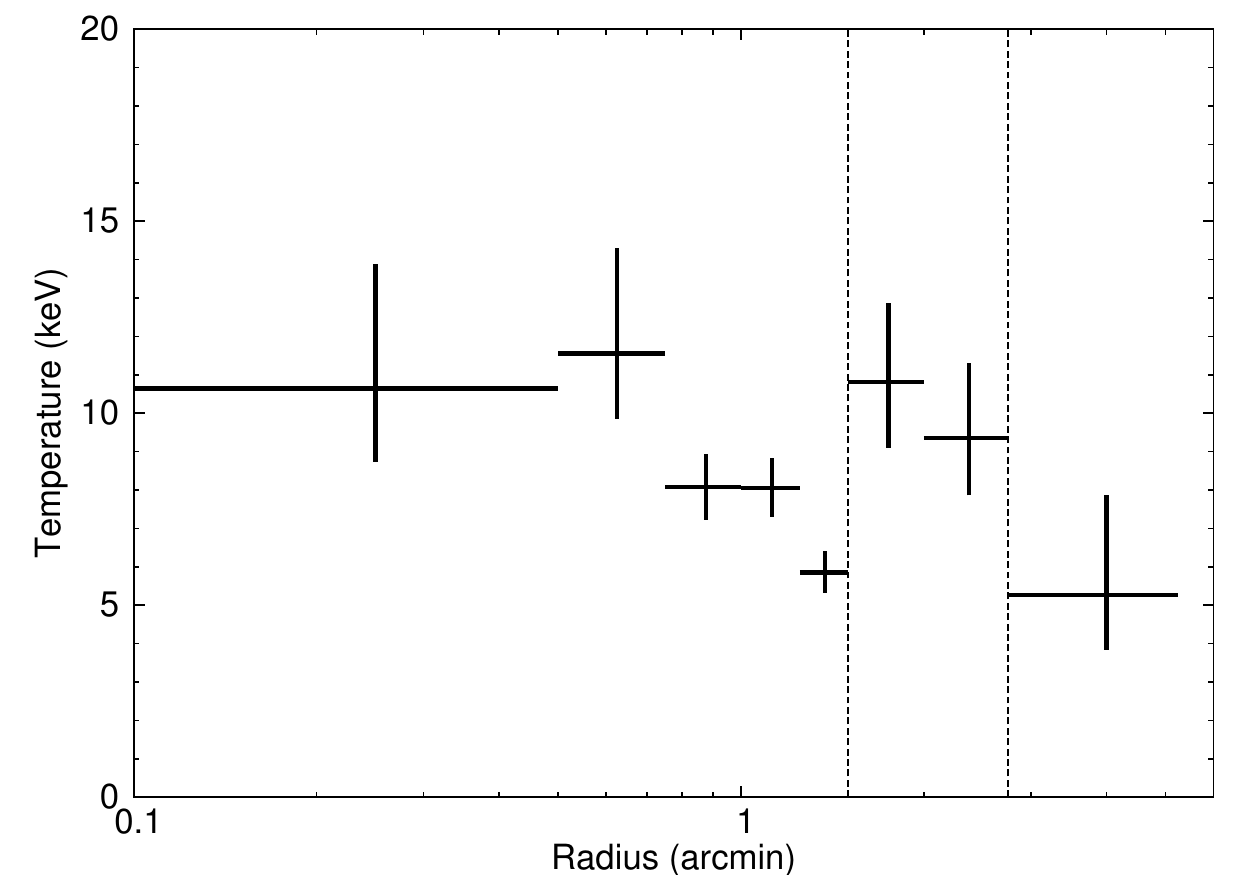}
\caption{Best-fit temperature profile for the gas in 3C 438. The region used to extract the temperature profile extends in the same direction as those used for the surface brightness profiles. The two jumps in temperature indicated by the lines are located at $1.5\arcmin$ and $3\arcmin$ from the center of the cluster.}
\label{fig:temperature}
\end{figure}

The second component required to compute the pressure jumps is the temperature profiles. To probe the temperature distribution of the hot gas across the surface brightness discontinuities, we build a temperature profile using circular wedges with a position angle of $180\degr-  220\degr$, as shown in Figure \ref{fig:temp_regions}. Note that these regions are similar to those used to extract the density profiles, except for the broader applied bins. In addition, we included a larger $2.5\arcmin \times 5\arcmin$ rectangular region covering the outskirts of the cluster, which was extracted from the front illuminated detector CCD 6. This large region is intended to probe the large-scale gas temperature of the cluster. We fit the data with a single-temperature \textsc{APEC} model, where we fixed the redshift to $z=0.29$, the column density to $N_{\rm H} = 2.7 \times 10^{21} \ \rm{cm^{-2}}$, and metallicity to $0.4$ Solar, leaving the temperature and normalization as free parameters. The obtained temperature profile is shown in Figure \ref{fig:temperature}. The profile reveals two evident jumps: at $1.5\arcmin$ we detect a factor of   $1.85\pm 0.38$ temperature increase, while at $3 \arcmin$ the temperature drops by a factor of $1.77 \pm 0.73$.

\begin{table}
\begin{center}
\caption{Best fit parameter values from Proffit \label{tbl-2}}
\begin{tabular}{lclcc}
\tableline
\tableline
Parameter & Value & Error \\
\tableline
$\beta$   & 0.30 & Fixed\\
$\alpha_1$ & 1.06 & $0.057$ \\
$\alpha_2$ & 0.98 & $0.24$ \\
$r_c$ (arcmins) & 0.74 & $0.39$  \\
$r_{cut, inner}$ (arcmins) & 1.53 & $0.0047$ \\
$r_{cut, outer}$ (arcmins) & 3.14 & $1.0 \times 10^{-5}$ \\
$n_{0,cut}/n_{1,cut}$ & 2.33 & 0.15 \\
$n_{1,cut}/n_{2,cut}$ & 3.55 & 0.76 \\
\tableline
\end{tabular}
\end{center}
\end{table}

\begin{figure*}
\centering
\includegraphics[width=8cm]{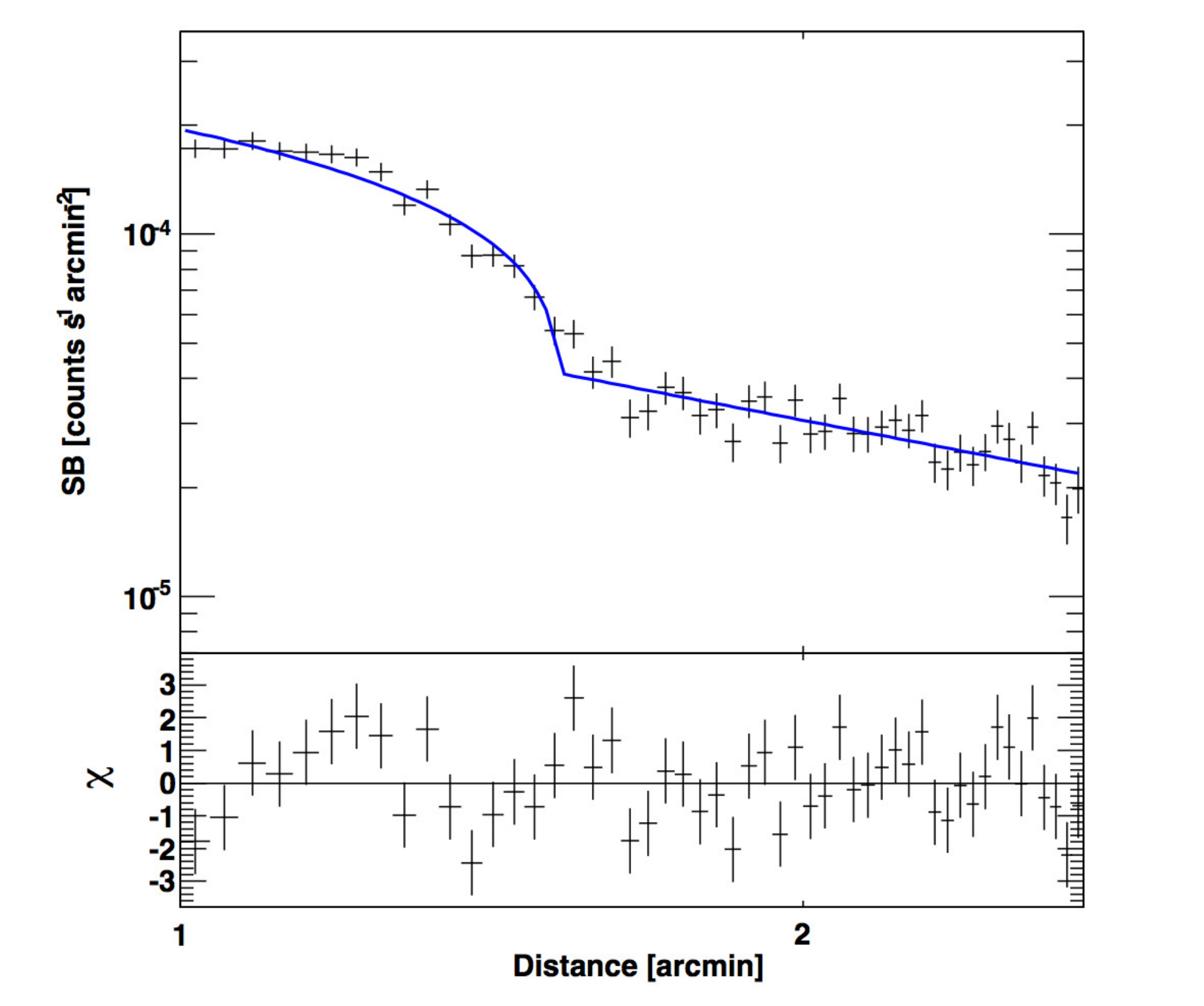}
\includegraphics[width=8cm]{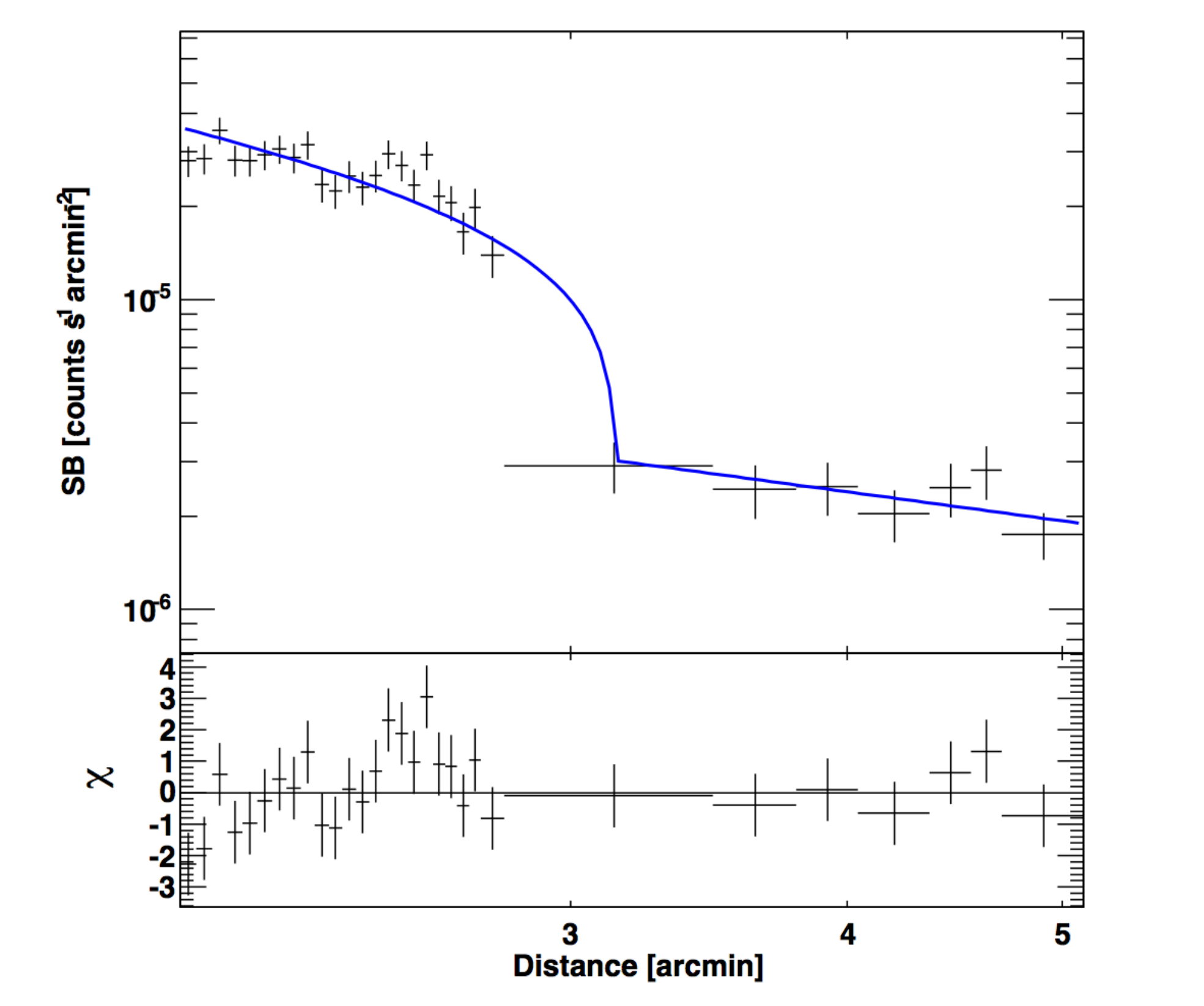}
\caption{The deprojected surface brightness profile for the inner jump (left) and outer jump (right). The best fits for the deprojected profiles are plotted and their residuals ($\chi $) are shown in the bottom panels of the graphs. The inner jump, shown in the left panel, was fitted using a $\beta$-model and a power law, while the outer jump, shown in the right panel, was fitted using two power laws. The best-fit model parameters are given in Table 2.}
\label{fig:density}
\end{figure*}

\subsection{Pressure Ratios and the Bulk Velocity}
Based on the magnitude of the density and temperature jumps at the inner and outer discontinuities, we derive the pressures as  $p = 1.9 n_e kT$ and compute the pressure ratios between the inside and outside of the jumps. Based on the above computed density and temperature jumps, we find that the pressure ratio is  $1.2 \pm 0.3$ at the inner jump, which indicates approximately continuous pressure within statistical uncertainties. However, at the position of the outer jump, we measure a pressure drop of $6.3 \pm 2.9$. 

Based on the large pressure jump at $3\arcmin$, we use the Rankine-Hugoniot jump conditions to find the velocity of the cluster. These give us:

\begin{equation}
\frac{P_2}{P_1}=\frac{2\gamma M_1^2 -\gamma +1}{\gamma + 1}
\label{equ:mach_pressure}
\end{equation}
\begin{equation}
\frac{\rho_2}{\rho_1}=\frac{(1+\gamma )M_1^2}{2+(\gamma - 1)M_1^2}
\label{equ:mach_density}
\end{equation}


\begin{figure}
\centering
\includegraphics[width=8cm]{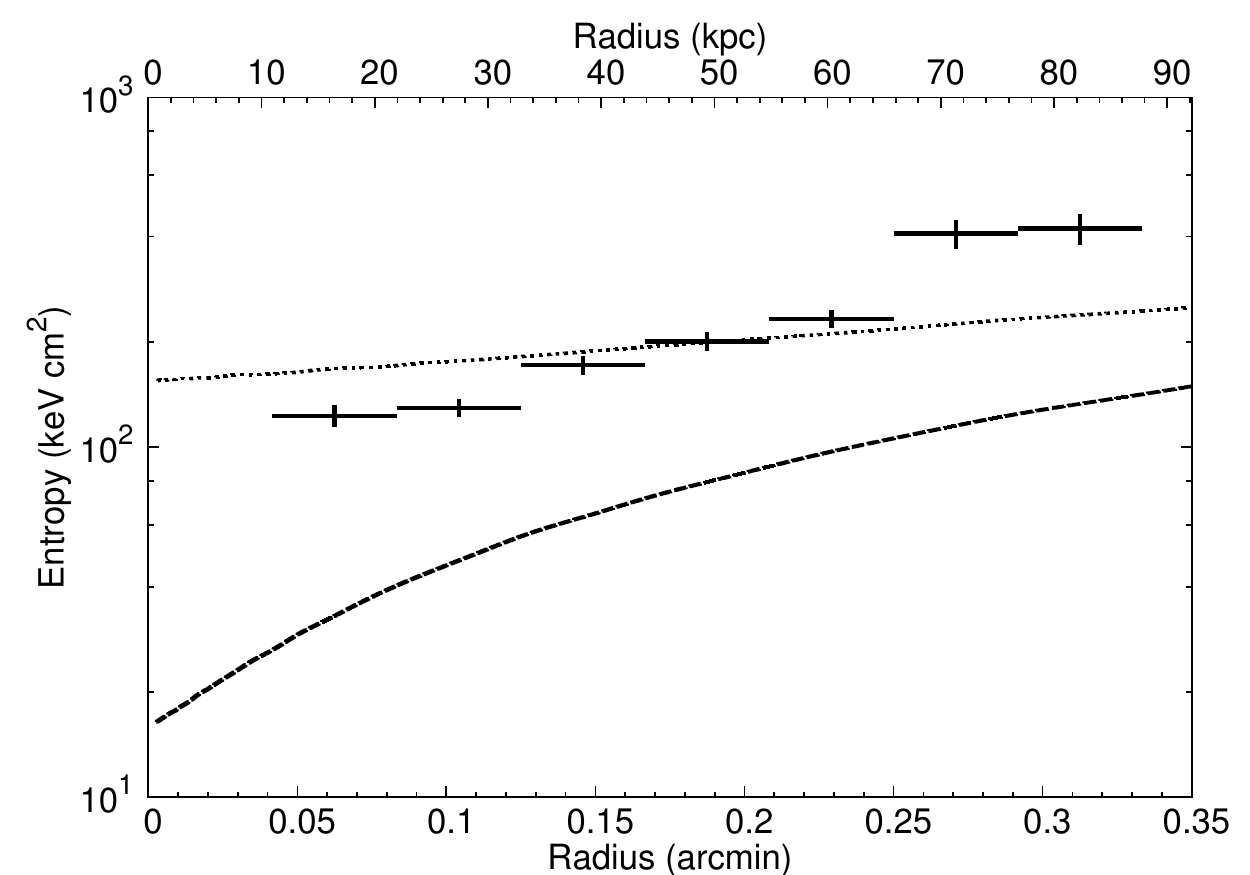}
\caption{Entropy as a function of distance from the center of the cluster core. The dashed and dotted lines represent the mean entropy profiles for cool core and non-cool core galaxy clusters respectively \citep{Cavagnolo2009}. The entropy profile of 3C 438 is similar to those observed in non-cool core clusters. In addition, the flatness of the curve near the center is also characteristic of non-cool core clusters.}
\label{fig:entropy}
\end{figure}


\begin{figure*}
\centering
\includegraphics[width=16cm]{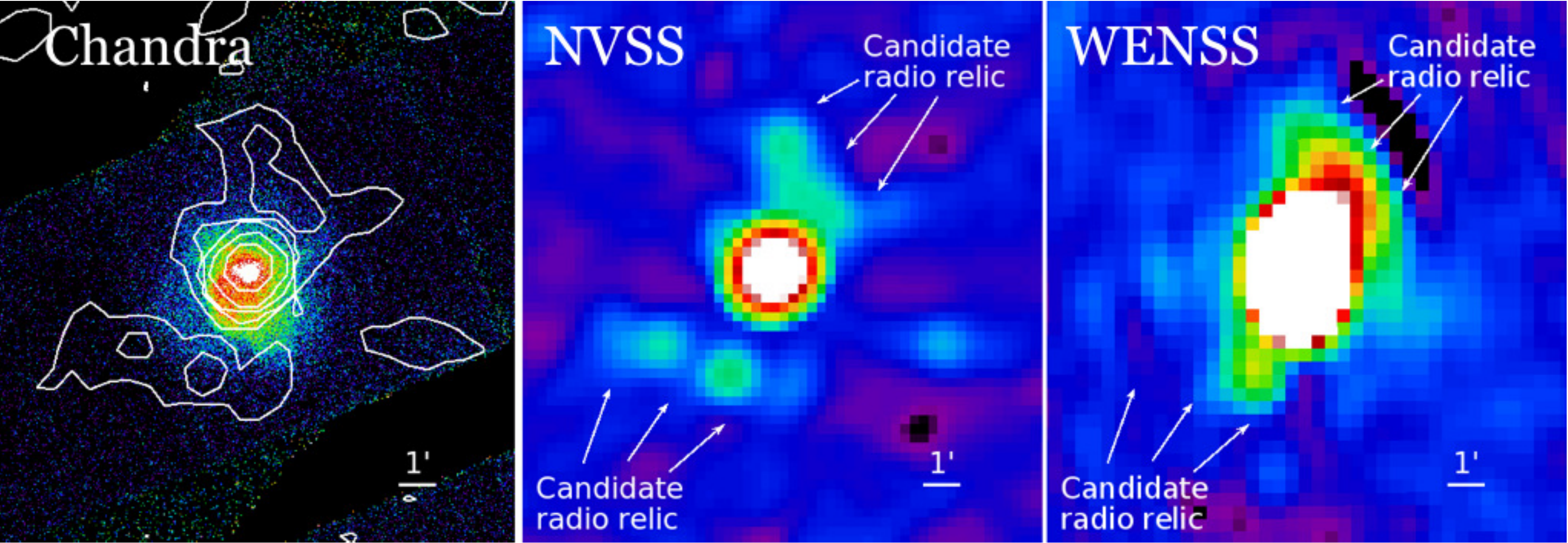}
\caption{Left: Large scale 0.5-4 keV band \textit{Chandra} X-ray image of the galaxy cluster around 3C 438. Overlaid are the $1.4$ GHz NVSS intensity contour levels, where the contour levels correspond to [0.001, 0.000654, 0.0377, 0.213, 1.2] mJy~beam$^{-1}$. The sharp outer surface brightness discontinuities at the eastern side of the cluster mark the position of the bow shock. Middle: 1.4 GHz NVSS image of the same region. The radio lobes associated with the radio galaxy, 3C 438, cannot be resolved. At large radii, extended radio emission appears to be present at the northwest and southeast of the cluster. Interestingly, the diffuse radio emission located at the southeast traces the shock front, which may hint that the diffuse radio emission is associated with radio relics, which are often observed around merging galaxy clusters. Right: 325 MHz WENSS radio image of the same region. Again, the arc-like features most clearly seen towards the northwest of the cluster suggest the possible presence of radio relics.}
\label{fig:nvss}
\end{figure*}

Using Equation \ref{equ:mach_pressure} and our value for the pressure jump, we find that the 3C 438 cluster is moving at Mach $M = 2.3\pm 0.5$. When using Equation \ref{equ:mach_density} to constrain the Mach number, we obtain $M = 4.9\pm 4.1$, which is consistent with that obtained from Equation \ref{equ:mach_pressure}. However, we note that due to the proximity of our density ratio to a critical point in the equation, located at $\rho_2 / \rho_1 = 4$, the latter Mach number has large uncertainty. The sound speed in a 5 keV plasma corresponds to $c_s=\sqrt{(\gamma kT)/(\mu m_H})=1130 \ \rm{km \ s^{-1}}$ by adopting  $\gamma=5/3$ and $\mu=0.62$. Thus, we find that the sub-cluster is moving supersonically with a velocity of $2600\pm 565$ km/s within the intracluster medium. This value should be considered as a lower limit since we do not account for the surrounding cool gas that is inherently present in any cross-section of the cluster. This causes the temperature measurement, pressure ratio and, therefore, the bulk velocity to be an underestimate.

\subsection{Major cluster merger}
The presence of sharp surface brightness discontinuities, combined with the observed pressure ratios across the edges, indicates that the inner discontinuity is a cold front \citep{Vikhlinin2001,Markevitch2007}. However, the drop in temperature by a factor of $1.77 \pm 0.73$ at $3\arcmin$ and the drop in density by a factor of $3.55 \pm 0.76$ imply a factor of $6.3 \pm 2.9$ drop in gas pressure, which in turn indicates that the outer discontinuity is a shock front. The temperature patterns, as well as the observed cold front and shock fronts in the cluster, are characteristic of mergers between two roughly equal mass sub-clusters. Thus, the galaxy cluster associated with 3C 438 is likely undergoing a major merger.
Moreover, the bulk motion of the gas in the cluster, having a Mach number of $M = 2.3\pm 0.5$, signifies a supersonic velocity within the intracluster medium. Along with the cone-like shape of the outer discontinuity (see Figure \ref{fig:outlines}), which has similarities to the features of the Bullet cluster, this high velocity suggests that the outer front is a bow shock.

An important feature to note is the fact that only one core is visible in the cluster. Since there is evidence that the 3C 438 cluster is undergoing a major merger, the appearance of only one cluster core hints that one of the clusters may be a non-cool core cluster. Non-cool core clusters have no notable drop in temperature towards the center of the cluster, causing the surface brightness to be more evenly spread \citep{DeGrandi2002}, which would explain the lack of a second nucleus. This is consistent with the scenario in which at least one of the merging sub-clusters is a non-cool core cluster.

In addition, we derived the (pseudo) entropy profile of the sub-cluster using
\begin{equation}
K=\frac{kT}{\mu m_H\rho^{2/3}}
\end{equation}
where $\mu$ is the mean molecular weight of the gas. We compare the entropy profile of the sub-cluster (Figure \ref{fig:entropy}) with the mean entropy profile of cool core clusters and non-cool core clusters \citep{Cavagnolo2009}. The entropy profile of 3C 438 is in good agreement with those observed for non-cool core clusters. Specifically, the entropy profile demonstrates a slow increase in entropy from the center, which is characteristic of non-cool core clusters. This hints that the second sub-cluster may also be a non-cool core cluster. Thus, these results imply that the galaxy cluster associated with 3C 438 may be undergoing a major merger between two non-cool core clusters.

\section{Discussion}
\label{sec:discussion}

In spite of its inclusion in the 3C catalog of radio galaxies \citep{Bennett1962}, 3C\,438 has been relatively poorly studied because of its low Galactic latitude. Low-resolution radio imaging was presented by \cite{Leahy1991}, and high-resolution radio imaging showing the unusual strong jets and weak hotspots of the radio source was carried out by \cite{Hardcastle1997} and \cite{Treichel2001}. \cite{mjh08} made maps of the central radio galaxy at 90 GHz detecting the nucleus and hot spots, and the broad band emission from the active nucleus has been well-studied as part of several radio galaxy surveys \citep{mingo14,Massaro15}. Most recently \cite{Harwood2015} have carried out a study of the spectral and dynamical ages of the radio galaxy, finding that it began to be active between 3 and 9 Myr ago. Thus, it seems unlikely that the current radio source has any relationship to the large-scale disturbances seen in the cluster, as argued on energetic grounds by \cite{Kraft2007}. Any shock features driven by the currently active radio galaxy would be expected to be seen very close to the radio lobes.

The linear scale of the radio galaxy 3C 438 is $\sim45$ kpc \citep{Kraft2007}, hence it is about an order of magnitude smaller than the surface brightness discontinuities presented in this work. To probe the radio emission on scales comparable with the bow shock, we present the $1.4$ GHz NRAO VLA Sky Survey (NVSS) and the 325 MHz Westerbork North Sky Survey (WENSS) images along with the $0.5-4$ keV band X-ray image in Figure \ref{fig:nvss}. Due to the low resolution of the NVSS image, the structure of the radio lobes of the central galaxy are not resolved. However, the large-scale radio image reveals enhanced emission towards the southeast and northwest of the galaxy cluster. The origin of these features is not clear. Although the position and symmetric nature of these features hint that they may be associated with the galaxy cluster, we cannot rule out the possibility that the emission originates from unresolved sources. To differentiate between these scenarios, and reveal whether these features may be radio relics typically observed around merging galaxy clusters \citep[e.g.][]{vanWeeren2010}, we would require more sensitive and higher resolution radio maps.

We plan to follow up this X-ray study with additional radio and optical observations. The presence of a powerful merger shock and the existing low-resolution radio observations suggest that this cluster may host a radio relic. To our knowledge, no moderate resolution radio observations at low frequency have been made of this cluster. Radio relics are often associated with strong merger shocks, and direct measurement of the shock strength in the X-ray band is critical for elucidating the underlying physics of non-thermal particle acceleration in the relics \citep{vanWeeren2010}. Additionally, this system is one of the few known to be in the early stages of a supersonic merger of two roughly equal mass clusters. Other examples include the Bullet cluster and Abell 520 \citep{Markevitch2002,Markevitch2007}. A survey of the galaxies in the two clusters could provide additional clues to better constrain the dynamics of the system. In particular, it would be interesting to know if the galaxies of the infalling cluster are offset from the gas in a similar manner to what has been observed with the Bullet cluster.

\section{Conclusions}
\label{sec:conclusions}

In this paper we utilized \textit{Chandra} X-ray observations to characterize the galaxy cluster associated with radio galaxy 3C 438 and to determine the origin of the surface brightness discontinuities. Our main results are:

\begin{itemize}
\item We identified two sharp surface brightness edges on the eastern side of the cluster. The inner jump is located $1.5\arcmin$, or 400 kpc, from the center of the cluster, while the outer jump is located $3\arcmin$, or 800 kpc, from the center of the cluster.

\item We derived density and temperature profiles, from which we were able to calculate the pressure jumps across the discontinuities, which showed the presence of a cold front at the inner discontinuity and a shock front at the outer discontinuity.

\item Based on the density, temperature, and pressure jumps, we concluded that there is a bow shock at the outer discontinuity and that the gas is moving with a bulk velocity of at least $M = 2.3 \pm 0.5$, or $2600\pm 565$ km/s.
\vspace{.18cm}
\item Finally, only one cluster core was evident in the merger, which was found to be a non-cool core. Thus, the observational data are consistent with a picture in which the 3C 438 cluster is undergoing a merger between two roughly equal mass, non-cool core clusters. \\ \\

\end{itemize}

\vspace{1cm}

Acknowledgments. We thank the referee for the constructive comments that have helped us to improve the paper. This research has made use of Chandra archival data provided by the Chandra X-ray Center. The publication makes use of software provided by the Chandra X-ray Center (CXC) in the application package CIAO. 
We have also used an NVSS observation from the National Radio Astronomy Observatory, which is a facility of the National Science Foundation operated under cooperative agreement by Associated Universities, Inc. 
Additionally, we have made use of the WSRT on the Web Archive. The Westerbork Synthesis Radio Telescope is operated by the Netherlands Institute for Radio Astronomy ASTRON, with support of NWO. 
Finally, A.B., W.R.F., C. J. are supported by the Smithsonian Institution.


\begin{thebibliography}{}


\bibitem[\protect\citeauthoryear{Planck(2015)}{Ade \etal}{2015}]{Planck2015} Ade P.~A.~R., Aghanim N., Arg{\"u}eso F., Arnaud M., Ashdown M., Aumont J., Baccigalupi C., Banday A.~J., Barreiro R.~B. \& et al., 2015, ArXiv e-prints, 1507.02058

\bibitem[\protect\citeauthoryear{Anders1989}
 {Anders \& Grevesse}{1989}]{Anders2006}
Anders E., Grevesse N., 1989, Geochimica et Cosmochimica Acta, 53, 197

\bibitem[\protect\citeauthoryear{Ascasibar2006}
 {Ascasibar \& Markevitch}{2006}]{Ascasibar2006}
Ascasibar Y., Markevitch M., 2006, ApJ, 650, 102

\bibitem[\protect\citeauthoryear{Bennett1962}{Bennett}{1962}]{Bennett1962} Bennett A.~S., 1962 \memras, 68, 163

\bibitem[\protect\citeauthoryear{Bogdan2008}
 {Bogdan \& Gilfanov}{2008}]{Bogdan2008}
Bogdan A., Gilfanov M., 2008, Monthly Notices of the Royal Astronomical Society, 388, 56

\bibitem[\protect\citeauthoryear{Bogdan2011}
 {Bogdan et al.}{2011}]{Bogdan2011}
Bogdan A., Kraft, Ralph P., Forman W. R., Jones C., Randall, S. W., Sun M., O'Dea C. P., Churazov E., Baum S. A., 2011, ApJ, 743, 11

\bibitem[Cavagnolo(2009)]{Cavagnolo2009} Cavagnolo K. W., Donahue M., Voit G.~M. Sun, M., ApJ, 182, 12

\bibitem[Close et al.(2013)]{Close2013} Close J.~L., Pittard J.~M., Hartquist T.~W., Falle S.~A.~E.~G., 2013, \mnras, 436, 3021 

\bibitem[\protect\citeauthoryear{DeGrandi2002}{De Grandi \& Molendi}{2002}]{DeGrandi2002} De Grandi S., Molendi S., 2002, \apj, 567, 163-177

\bibitem[\protect\citeauthoryear{Eckert2011}
 {Eckert, Molendi \& Paltani}{2011}]{Eckert2011}
Eckert D., Molendi S., Paltani S., 2011, Astronomy and Astrophysics, 526, 15

\bibitem[\protect\citeauthoryear{Fruscione2006}
 {Fruscione et al.}{2006}]{Fruscione2006}
Fruscione A. \etal, 2006, Observatory Operations: Strategies, Processes, and Systems, 6270, 95


\bibitem[\protect\citeauthoryear{Guver \& Ozel}{Guver \& Ozel}{2009}]{Guver2009} G{\"u}ver T., {\"O}zel F., 2009, \mnras, 400, 2050-2053


\bibitem[\protect\citeauthoryear{Hardcastle1997}{Hardcastle \etal}{1997}]{Hardcastle1997} Hardcastle M. J., Alexander P., Pooley G.~G., Riley J.~M., 1997, \mnras, 288, 859-890

\bibitem[\protect\citeauthoryear{Hardcastle \& Looney}{Hardcastle \& Looney}{2008}]{mjh08} Hardcastle M. J., Looney L. W., 2008, \mnras, 388, 176

\bibitem[\protect\citeauthoryear{Harwood2015}{Harwood \etal}{2015}]{Harwood2015} Harwood J.~J., Hardcastle M.~J., Croston J.~H., 2015, \mnras, 454, 3403-3422

\bibitem[\protect\citeauthoryear{Hickox \& Markevitch}
 {Hickox \& Markevitch}{2006}]{hickox2006}
Hickox R. C., Markevitch M., 2006, ApJ, 645, 95


\bibitem[\protect\citeauthoryear{Kraft2007}
 {Kraft et al.}{2007}]{Kraft2007}
Kraft R. P., Forman W. R., Hardcastle M. J., Jones C., Nulsen P. E. J., 2007, ApJ, 664:L83-L86

\bibitem[\protect\citeauthoryear{Kravtsov2012}
 {Kravtsov \& Borgani}{2012}]{Kravtsov2012}
Kravtsov, A. V., Borgani S., 2012, Annual Review of Astronomy and Astrophysics, 50, 353

\bibitem[\protect\citeauthoryear{Leahy1991}{Leahy \& Perley}{1991}]{Leahy1991} Leahy J.~P., Perley R.~A., 1991, \aj, 102, 537-561

\bibitem[\protect\citeauthoryear{Markevitch2002}
 {Markevitch \etal}{2002}]{Markevitch2002}
Markevitch M., Gonzalez A. H., David L., Vikhlinin A., Murray S., Forman W., Jones C., Tucker W., 2002, ApJ, 567:L27-L31

\bibitem[\protect\citeauthoryear{Markevitch2007}
{Markevitch \& Vikhlinin}{2007}]{Markevitch2007}
  Markevitch M., Vikhlinin A., 2007, Phys.Rep., 443
  
  \bibitem[\protect\citeauthoryear{Massaro15}{Massaro \etal}{2015}]{Massaro15} Massaro F. \etal, 2015, ApJS, 220, 5
  
  \bibitem[\protect\citeauthoryear{Mingo \etal.}{Mingo \etal}{2014}]{mingo14} Mingo B., Hardcastle M. J., Croston J. H., Dicken D., Evans D. A., Morganti R., Tadhunter C., 2014, MNRAS, 440, 269.\

\bibitem[Nulsen(1982)]{Nulsen1982} Nulsen, P.~E.~J., 1982, \mnras, 198, 1007 

\bibitem[\protect\citeauthoryear{Treichel2001}{Treichel \etal}{2001}]{Treichel2001} Treichel K., Rudnick L., Hardcastle M.~J., Leahy J.~P., 2001, \apj, 561, 691-702

\bibitem[\protect\citeauthoryear{vanWeeren2010}{van Weeren \etal}{2010}]{vanWeeren2010} van Weeren R. J., R{\"o}ttgering H. J. A., Br{\"u}ggen M., Hoeft M., 2010, Science, 330, 347

\bibitem[\protect\citeauthoryear{Vikhlinin2001}
 {Vikhlinin, Markevitch \& Murray}{2001}]{Vikhlinin2001}
Vikhlinin A., Markevitch M., Murray S. S., 2001, ApJ, 551, 160

\end{thebibliography}

\end{document}